\documentclass[reqno,a4paper]{amsart}
\usepackage{amssymb,stmaryrd}
\usepackage{amsfonts}
\usepackage{amstext}
\usepackage{color}
\usepackage{graphicx}
\usepackage{subcaption}
\usepackage{hyperref}
\usepackage{amsmath}
\usepackage{booktabs}
\usepackage{placeins}

\newcommand{\extd}{\mathrm{d}}

\newcommand{\del}{\partial}

\renewcommand{\imath}{{\mathfrak{i}}}
\renewcommand{\div}{{\mathrm{div}}}

\begin{document}

\title{Geometric Aharonov--Bohm phase effect around a black hole}

\author[Kaushlendra Kumar and Shahn Majid]{\large Kaushlendra Kumar and Shahn Majid \\ \ \\ 
  School of Mathematical Sciences\\ 
  Queen Mary University of London\\ 
  Mile End Road, London E1 4NS, UK}

\thanks{
{\it Authors to whom correspondence should be addressed:}  kaushlendra.kumar@qmul.ac.uk and s.majid@qmul.ac.uk}

\thanks{{\it Funding:} KK was supported by a DFG project grant 515782239 and SM by a Leverhulme Trust project grant RPG-2024-177}
\date{June 2025 Ver 1.2}



\date{\today}

\begingroup
\let\MakeUppercase\relax % Disables \MakeUppercase for authors
\maketitle
\endgroup

\author{Kaushlendra Kumar and Shahn Majid}

\begin{abstract}In recent work, we have explored the novel possibility that a spacetime  geodesic density flow  in GR could be upgraded to the flow of an amplitude $\psi$ with density $|\psi|^2$. We now show how exactly the relevant amplitude and velocity flow equations arise in an eikonal approximation of Stueckelberg's proper-time quantum mechanics or Klein-Gordon flow. We also show how the divergence of the velocity field recovers the Raychaudhuri equations for relativistic fluids. In this setting, we demonstrate an Aharonov--Bohm type effect for the phase of $\psi$ in motion approaching a black-hole. 
\end{abstract}

\section{Introduction}

In recent work there has been renewed interest in wave-functions $\varphi$ on spacetime evolving with respect to an external proper time $s$ as a `Klein-Gordon flow'\cite{BegMa:gen, BegMa:flrw},
\begin{equation}\label{KGflow}  -\imath\hbar {\del\varphi\over\del s} = {\hbar^2\over 2 m}\square \varphi\end{equation}
where one can also add an external potential in analogy with quantum mechanics. In fact, this proposal of `proper-time quantum mechanics' turns out to have a long history that traces back to E.C.G. Stueckelberg \cite[Eqn. (5.2)]{Stu} in 1942 with $s=\pm m\lambda$. Later works \cite{Coo, HorPir} followed up this idea at least in flat spacetime and it has continued to surface from time to time. Until now, however, the physical meaning of $s$ has remained unclear and is one of the things we will address. The starting point already in \cite{Stu} is the geodesic motion of a single particle with proper time $s$. This was then extrapolated to a Heisenberg picture flow with the same $s$ and evolution in the Heisenberg algebra in the case of flat spacetime. Recently, the Heisenberg picture was extended in \cite{BegMa:gen} using noncommutative geometry to a general spacetimes as a theory of generally covariant quantum mechanics. Here, the evolution is in the algebra of differential operators (which in any local coordinate chart looks like the usual Heisenberg algebra). The corresponding Schr\"odinger picture of a flow on wavefunctions remains (\ref{KGflow}) for the wave operator provided by the Levi-Civita connection. The earlier flat space version in \cite{BegMa:geo} also includes minimal coupling to an electromagnetic field in the spirit of \cite{Stu}. 

The precise construction in \cite{BegMa:geo,BegMa:gen} is as a quantum-geometric geodesic flow.   In noncommutative geometry, as there are typically no points, there cannot be single geodesics either, but one can work in analogy with (relativistic) fluid mechanics by replacing a point particle by a density $\rho$. In the general case\cite{Beg:geo}, $\rho$ is a positive element of a $*$-algebra $A$ which could be noncommutative (such as the Heisenberg algebra or the algebra of differential operators), which one can specialise to ones of the form $\rho=\psi^*\psi$ where $\psi$ in the classical case would be an amplitude with $\rho=|\psi|^2$. Although the classical case $A=C^\infty(M)$ on spacetime $M$ was motivation, it turns out to be interesting in its own right and was studied around a black hole in   \cite{KumarMajid2025} as a novel possibility coming out of noncommutative geometry. Such `classical quantum geodesic flows' have everything classical so that one  can use the regular tools of GR and relativistic fluid mechanics, {\em except} that we consider an underlying {\em geodesic amplitude flow} for a wave-function $\psi$ rather than a density. Here, $\psi$ evolves according to
\begin{equation} 
  \dot\psi+X(\psi)+\frac12\,\div(X)\psi=0,
  \label{eq:flowpsi}
\end{equation}
where the dot denotes \(\partial/\partial s\) and $X$ is an auxiliary timelike vector field that is subject to the {\em geodesic velocity flow}
\begin{equation}
  \dot X+\nabla_X X=0.
  \label{eq:flowX}
\end{equation}
The amplitude flow then implies the usual density flow 
\begin{equation} 
  \dot\rho+X(\rho)+\rho\,\div(X)=0
  \label{eq:flowrho}
\end{equation}
and if $\nabla$ is the Levi-Civita connection and we use the usual $\sqrt{-g}$-weighted measure of integration then the integral of $\rho$, i.e. the  $L^2$-norm of $\psi$, is constant.  It is shown numerically in \cite{KumarMajid2025} that this amplitude hypothesis is testable in that two density bumps colliding merge to a bigger density bump but two amplitude bumps of opposite sign colliding merge to a dipole with a bipartite density profile of two bumps very close to each other but strictly separated. In the present sequel, we extend this to include a gravitational Aharonov-Bohm effect on the phase of $\psi$.

We first show, in Section~\ref{sec:wkb}, that the above wavefunction hypothesis for $\psi$ can in fact be derived from Stueckelberg's proper-time quantum mechanics/KG flow  as an eikonal/WKB approximation for $\varphi$. Hence, we have  a more precise meaning of $s$ in (\ref{KGflow}) in this approximation. Although still theoretical in that one needs to set up the initial state over all spacetime and ask what it is at time $s$, the meaning of $s$ is that each particle in the initial density moves along a geodesic by its own proper time $s$. So $s$ is not the proper time of any one particle but of all particles in some sense, except that we do not work with actual particles but with densities. We have called it the {\em collective proper time} for this reason and confirmed the interpretation against an actual bunch of geodesics smoothed with Gaussian interpolation in \cite{KumarMajid2025}. Meanwhile $X$ plays the same role as the fluid 4-velocity in relativistic fluid mechanics and we show in Section~\ref{sec:ray} how an equation for the convective derivative of ${\rm div}(X)$ involving the Ricci tensor in \cite{BegMa:cur} indeed recovers the Raychaudhuri equations for relativistic fluids.

In this paper we then study how the phase of complex amplitude $\psi$ is transported independently of the modulus. If \(\psi=\sqrt{\rho}e^{i\vartheta}\) is a smooth polar decomposition, the additional information in \eqref{eq:flowpsi} beyond the evolution of the density 
amounts to
\begin{equation}
  \dot\vartheta+X(\vartheta)=0 .
  \label{eq:flowphase}
\end{equation}

The equations solved in \cite{KumarMajid2025}  were, however, only involved radius-time coordinates (or more precisely Kruskal-Szekeres $U-V$ coordinates) with no angular dependence. As a result, we were not placed to demonstrate an Aharonov--Bohm (AB) effect which should be a further outcome of the wave-function hypothesis. 
To achieve this, we now go one step higher and work in the axially symmetric \((U,V,\theta)\) system to ask how the same black-hole geodesic flow transports an initially prepared angular phase. The central theoretical result is the reduced phase law
\begin{equation}
  \varphi_k(s,x) = -k\int_0^s X^\theta(\tau,\gamma(\tau))\,d\tau ,
  \label{eq:phase-law}
\end{equation}
where \(\gamma\) is the backward integral curve of the reduced geodesic velocity field \(X\), ending at the observation point \(x\) at flow time \(s\) and $k$ is the angular mode number for the $\theta$ variable of the spherical polar coordinates of each sphere at a given $r,t$. This phase law~\eqref{eq:phase-law} amounts to a geometric AB-type mechanism, but coming from the geometry rather than an electromagnetic solenoid or holonomy as would normally be the case. The mode number \(k\) plays the role of a charge, \(X^\theta\) plays the role of the connection component, and the observable consequence is a shift of an interference profile. We then proceed to numerically illustrate the phase law in two complementary ways. The direct reduced phase extracted from the solved wave-function gives the sharpest internal validation of the phase law. Secondly, a \(UV\)-integrated interference ratio \(\bar R_k(s,\theta)\), built from coherent superpositions, along wave-packet centroid ($\psi^2$-weighted packet centre location) provides a more physical intensity-level reading.  

In terms of organisation of this part of the paper, Section~\ref{sec:setup} sets up the 4D  flow equations around a black-hole and their axial reduction, along with Ehrenfest identities for the centroid. Section~\ref{sec:phaseframework} constructs  phase-carrying wave-packets, derives the exact solution and the above \(k\)-linear reduced phase law, and formulates the interference observable. Section~\ref{sec:numerics} provides the numerical validation: packet transport, direct reduced phase extraction, and the intensity-level phase extraction. 

We note that geometric and gravitational phase effects are already known in some other contexts. Berry's phase~\cite{Berry} concerns adiabatic transport in parameter space, while Colella et al.'s neutron experiment~\cite{ColellaOverhauserWerner} and the analyses of Anandan~\cite{Anandan} and Stodolsky~\cite{Stodolsky} concern gravitational or inertial phases in matter-wave interference. Closer to the present work are notions of an AB effect in a curved space setting. Early weak-field and global-spacetime versions include Papini's~\cite{Papini} gravitomagnetic matter-wave phase, Dowker's~\cite{Dowker} construction based on spacetime with nontrivial global topology, Stachel's~\cite{Stachel} globally stationary but locally static space-times, and the spin-connection/mass-current Dirac analogue of Lawrence et al.~\cite{LawrenceLeiterSzamosi}. More recent matter-wave versions by Hohensee et al.~\cite{Hohensee} and Overstreet et al.~\cite{Overstreet} concern gravitational-potential/redshift phases in force-free or near force-free interferometric settings while Jusufi et al.~\cite{Jusufi} use gravitational AB phases as a probe of Kaluza--Klein/Yukawa modified-gravity corrections. Our mechanism is novel and different from all of these examples and settings as it pertains to the flow of geodesics in an amplitude form of a wave-function on spacetime with the role of background field played by the angular component \(X^\theta\) of the geodesic velocity field.

\section{Links to Stueckelberg theory and the Raychaudhuri equation}\label{sec:pre}

This section provides the theoretical underpinning of classical quantum geodesics. We work on a pseudo-Riemannian manifold with metric $g$. All of the geometry is real as usual, only $\psi$ and $\varphi$ are complex. Given a vector field $X$, any scalar \(f\) or vector field \(Y\) has, as in fluid mechanics, a  {\em convective derivative} relative to it,
\[
  \frac{D}{Ds}\Big|_X f=\dot f+X(f),
  \qquad
  \frac{D}{Ds}\Big|_X Y=\dot Y+\nabla_XY .
\]
Then the geodesic velocity equation is ${D\over Ds}|_X X=0$ and similarly the length of \(X\) obeys\cite{BegMa:cur}
\begin{equation}
  \frac{D}{Ds}\Big|_X |X|^2=0.
  \end{equation}
 This means that if we initially set $|X|=-1$ (say) for a constant length timelike field then this feature will be preserved during the evolution. We will suppress all aspects of the underlying noncommutative geometry in favour of the equations in the introduction and the usual tools of GR.

\subsection{Derivation of geodesic flow from the Klein-Gordon flow}
\label{sec:wkb}

We demonstrate here how to obtain the flow system (\ref{eq:flowX}-\ref{eq:flowpsi}) as the leading complex-envelope semiclassical system associated with the Klein--Gordon flow (\ref{KGflow}). This is also needed to fix the distinction between the fast eikonal phase and the slow envelope phase used in the geometric-phase analysis in the main part of the paper. Thus, we use the complex-envelope WKB/eikonal ansatz
\begin{equation}
  \varphi(s,x)=\psi(s,x)e^{iS(s,x)/\hbar},
  \label{eq:envelopeansatz}
\end{equation}
where $S$ is a real rapidly varying phase and $\psi$ is a slowly varying \emph{complex} envelope. This eikonal phase $S$ should not be confused with phase $\vartheta$ of the slow envelope $\psi$ which gets convectively transported by \eqref{eq:flowphase}.

Since $\varphi$ is a scalar, one simply has
\begin{equation}
  \dot\varphi=e^{iS/\hbar}\left(\dot\psi+\frac{i}{\hbar}\psi\,\dot S\right),\quad \partial_\mu\varphi
  =e^{iS/\hbar}\left(\del_\mu\psi+\frac{i}{\hbar}\psi\del_\mu S\right),
  \label{eq:firstderPhi}
\end{equation}
where a second covariant derivative on the second part gives
\begin{equation}
  \Box\varphi
  =e^{iS/\hbar}\left(
    \Box\psi
    +\frac{2i}{\hbar}(\del^\mu S)\del_\mu\psi
    +\frac{i}{\hbar}(\Box S)\psi
    -\frac{1}{\hbar^2}|\extd S|^2\psi
  \right),
  \label{eq:boxPhiWKB}
\end{equation}
where $|\extd S|^2=(\del^\mu S)\del_\mu S$. Substituting this result for the box operator on $\varphi$ along with the first part of \eqref{eq:firstderPhi} into
\eqref{KGflow} yields,
\begin{equation}
  \psi\left(\dot S+\frac{1}{2m}|\extd S|^2\right)
  =\frac{\hbar^2}{2m}\Box\psi
  +i\hbar\left(
    \dot\psi
    +\frac{1}{m}(\del^\mu S)\del_\mu\psi
    +\frac{1}{2m}(\Box S)\psi
  \right).
  \label{eq:masterWKBidentity}
\end{equation}
Thus, we have the leading ${\mathcal O}(1)$ term producing the Hamilton--Jacobi equation for the eikonal phase $S$:
\begin{equation}
  \dot S + \frac{1}{2m}|\extd S|^2 = 0.
  \label{eq:HJflow}
\end{equation}
We define the gradient covector field
\begin{equation}
  X_\mu:={1\over m} \del_\mu S  \label{eq:XfromS}
\end{equation}
and $X^\mu$ the corresponding vector field via the metric. If we differentiate \eqref{eq:HJflow}, we get
\begin{equation}
 \dot X_\lambda + \frac{1}{2} \del_\lambda(X^\mu X_\mu)=0.
  \label{eq:delsP}
\end{equation}
Now $m X_\mu=\del_\mu S$ is a gradient of a scalar, hence
\begin{equation}
  \nabla_\lambda X_\mu = \nabla_\mu X_\lambda,
  \label{eq:gradintegrability}
\end{equation}
and (\ref{eq:delsP}) becomes
\begin{equation}
  \dot X^\nu + X^\mu\nabla_\mu X^\nu = 0.
  \label{eq:velflowcovupper}
\end{equation}
This is precisely the geodesic velocity equation \eqref{eq:flowX}. Note that the theory of classical quantum geodesics is more general in that $X$ (for example the one displayed in Figure~\ref{fig:velocityfield}) does not need to be a gradient. However,  this provides a natural class arising from a Klein-Gordon flow.

Now moving on to the next order $\mathcal{O}(\hbar)$ in \eqref{eq:masterWKBidentity}, one obtains after imposing \eqref{eq:HJflow},
\begin{equation}
  \dot\psi + X^\mu\del_\mu\psi + \frac12 (\nabla_\mu X^\mu)\psi = 0,
  \label{eq:semiclassicalpsi}
\end{equation}
which is exactly the amplitude-flow equation \eqref{eq:psiPDE4d}. We can ignore the residual $\mathcal{O}(\hbar^2)$ term in this semiclassical analysis.
Related WKBJ derivations of the Hamilton--Jacobi equation, the continuity equation for the leading amplitude density, and amplitude/spin transport along the associated mechanical trajectories were given by Stachel and Plebański~\cite{StachelPlebanski}.

\subsection{Raychaudhuri equation for geodesic flow}
\label{sec:ray}

Here we demonstrate the Raychaudhuri identity behind the divergence term appearing in amplitude flow equation \eqref{eq:psiPDE}. Letting $\Theta:=\nabla_\mu X^\mu$, it was shown in \cite{BegMa:cur} for any classical quantum geodesic flow that
\begin{equation}
  \frac{D\Theta}{Ds} = -(\nabla_\mu X^\nu)(\nabla_\nu X^\mu) -R_{\mu\nu}X^\mu X^\nu.
  \label{app:ray-start}
\end{equation}
We also recall that timelike constant length vector fields stay such, and for our analysis we choose  \(X^\mu X_\mu=-1\). (This is again beyond the WKB/eikonal case, where it would imply that $X_\mu$ is constant in $s$.) To recover the standard form of Raychaudhuri equation, we introduce the spatial projector orthogonal to \(X\),
\begin{equation}
  h_{\mu\nu}=g_{\mu\nu}+X_\mu X_\nu,
  \qquad
  h^\mu{}_\nu=\delta^\mu{}_\nu+X^\mu X_\nu.
  \label{app:projector}
\end{equation}
It satisfies \(h_{\mu\nu}X^\nu=0\) and has trace \(h^\mu{}_\mu=3\).  The
projected deformation tensor is
\begin{equation}
  B_{\mu\nu}
  :=
  h_\mu{}^\alpha h_\nu{}^\beta \nabla_\alpha X_\beta .
  \label{app:Bdef}
\end{equation}
It is orthogonal to \(X\) in both indices and therefore acts as a \(3\times3\)
matrix on the local rest space perpendicular to the flow.

It is useful to keep track of the acceleration part,
\begin{equation}
  a_\nu:=X^\alpha\nabla_\alpha X_\nu,
  \label{app:accel}
\end{equation}
which using \(X^\nu\nabla_\mu X_\nu=0\) from the unit-speed condition allows one to express the deformation tensor as,
\begin{equation}
  B_{\mu\nu}=\nabla_\mu X_\nu+X_\mu a_\nu,
  \quad\implies\quad
  \nabla_\mu X_\nu=B_{\mu\nu}-X_\mu a_\nu .
  \label{app:deform-accel}
\end{equation}
In the quadratic term in \eqref{app:ray-start}, however,
the acceleration contribution drops out:
\begin{equation}
  (\nabla_\mu X^\nu)(\nabla_\nu X^\mu)=B_\mu{}^\nu B_\nu{}^\mu,
  \label{app:accel-drops}
\end{equation}
due to $X$-orthogonality, \(B_{\mu\nu}X^\nu=X^\mu B_{\mu\nu}=0\) and \(X^\mu a_\mu=0\).

We now decompose the spatial \(3\times3\) tensor \(B_{\mu\nu}\) into its trace, symmetric traceless, and antisymmetric parts:
\begin{equation}
  B_{\mu\nu}
  =
  \frac13\Theta h_{\mu\nu}
  +\sigma_{\mu\nu}
  +\omega_{\mu\nu},
  \label{app:Bsplit}
\end{equation}
where
\begin{equation}
  \Theta=B^\mu{}_\mu,
  \qquad
  \sigma_{\mu\nu}=B_{(\mu\nu)}-\frac13\Theta h_{\mu\nu},
  \qquad
  \omega_{\mu\nu}=B_{[\mu\nu]} .
  \label{app:sigma-omega}
\end{equation}
The tensor \(\sigma_{\mu\nu}\) is the shear and \(\omega_{\mu\nu}\) is the vorticity. Using \eqref{app:Bsplit} into \eqref{app:accel-drops} we obtain
\begin{equation}
  B_\mu{}^\nu B_\nu{}^\mu = \frac13\Theta^2 + \sigma_{\mu\nu}\sigma^{\mu\nu} - \omega_{\mu\nu}\omega^{\mu\nu}.
  \label{app:Bcontract}
\end{equation}
Substituting this into \eqref{app:ray-start} and using scalar conventions
\(\sigma^2=\frac12\sigma_{\mu\nu}\sigma^{\mu\nu}\) and
\(\omega^2=\frac12\omega_{\mu\nu}\omega^{\mu\nu}\), we obtain the known form of the Raychaudhuri equation, albeit with a convective derivative:
\begin{equation}
    \frac{D\Theta}{Ds} = -\frac13\Theta^2 - 2\sigma^2 + 2\omega^2 -R_{\mu\nu}X^\mu X^\nu.
    \label{app:ray-final}
\end{equation}

Moreover, in the Schwarzschild vacuum background \(R_{\mu\nu}=0\), and, assuming zero vorticity (current work however deals with non-zero vorticity velocity fields), leads us to
\begin{equation}
  \frac{D\Theta}{Ds} = -\frac13\Theta^2 -2\sigma^2 \leq 0,
  \label{app:ray-vacuum}
\end{equation}
meaning that the expansion scalar would not increase in time. We can solve this along the integral curve $\gamma^\mu(\tau)$ \eqref{eq:IC} where $\frac{D\Theta}{Ds}=\frac{\extd\Theta}{\extd\tau}$ to obtain the real amplitude factor in the exact solution \eqref{eq:exactsol},
\[
  \psi(s,x)
  =
  \psi(0,\gamma(0))
  \exp\!\left[-\frac12\int_0^s \Theta(\tau,\gamma(\tau))\,d\tau\right].
\]
This gives the geometric meaning of the real amplitude factor in the exact solution: positive expansion attenuates the envelope, while focusing enhances it. This factor is real and positive, and therefore does not affect the reduced phase. The phase shift is instead determined by the angular foot-point of the backward integral curve, equivalently by the integral of \(X^\theta\) in \eqref{eq:phase-law}.

\label{sec:setup}

%-------------------------------------------------------------------
\section{Flow equations in 4D Schwarzschild--Kruskal setting and their axial reduction}
\label{sec:setup}
%-------------------------------------------------------------------
This section recaps the geometry of Schwarzschild and Kruskal--Szekeres coordinates and expands the flow equations in the full 4D framework. These are then reduced via axial symmetry as needed for the later sections. We also cover Ehrenfest theorems which are part of the justification of the geodesic flow parameter $s$ as proper time. 

\subsection{Schwarzschild and Kruskal--Szekeres coordinates}

We begin with the Schwarzschild metric in standard coordinates $(t,r,\theta,\phi)$, with Schwarzschild radius $r_s=1$,
\begin{equation}
  \extd s^2 = -\Bigl(1-\frac1r\Bigr)\extd t^2
  + \Bigl(1-\frac1r\Bigr)^{-1}\extd r^2
  + r^2\bigl(\extd\theta^2+\sin^2\!\theta\,\extd\phi^2\bigr).
  \label{eq:metrictr}
\end{equation}
This form suffers from the well-known horizon singularity at $r=1$, which is nothing but a coordinate artefact.

\begin{figure}[t]
\centering
\includegraphics[width=0.6\textwidth]{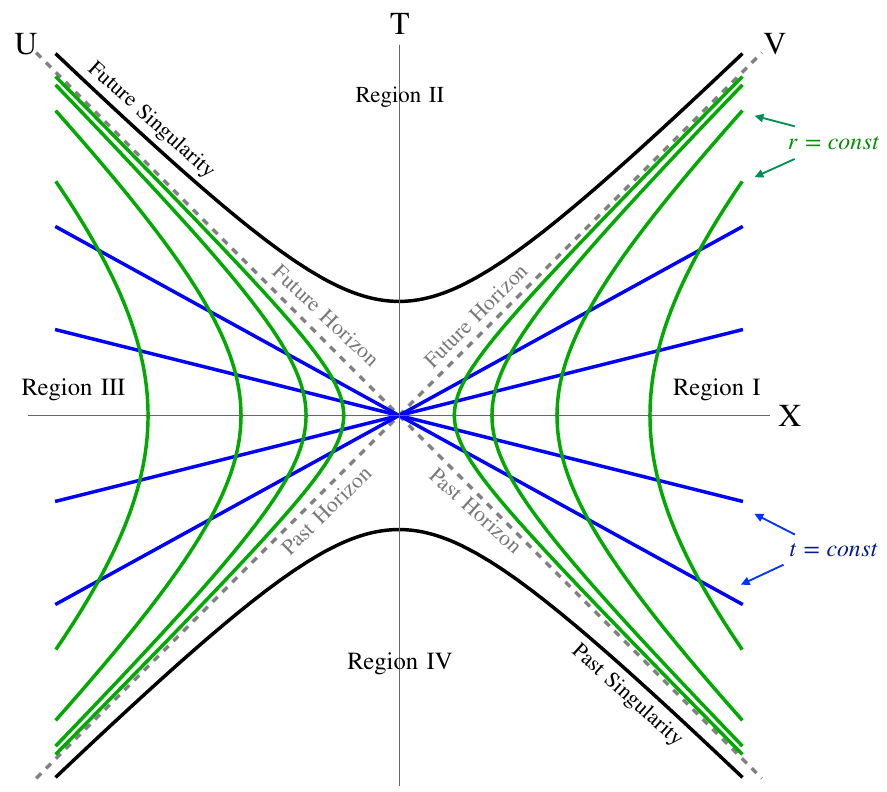}
\caption{Kruskal--Szekeres diagram with various axes and regions marked. The dashed grey lines are the horizons while the black hyperbolae are the singularities. Also, the constant Schwarzschild time/radius slices in the exterior are depicted in blue/green.}
\label{KSplot}
\end{figure}

To describe the geodesic velocity field (and possible wave packet motion) all the way inside horizon without issues, we work instead in Kruskal--Szekeres double-null coordinates $(U,V,\theta,\phi)$. The extended spacetime, in these coordinates, is divided into four regions separated by the hypersurfaces $U=V=0$: Region~I ($U<0$, $V>0$), the exterior; Region~II ($U>0$, $V>0$), the black-hole interior; Region~III ($U>0$, $V<0$), the second exterior (interpreted as parallel inaccessible region); and Region~IV ($U<0$, $V<0$), the white-hole interior (a time reversed black-hole region where nothing falls in). For our considerations here, the relevant spacetime sector is region-I where the future horizon is $\{U=0,\,V>0\}$ and the past horizon is $\{V=0,\,U<0\}$, as depicted in Figure~\ref{KSplot}.

More concretely, this is easily seen by introducing the tortoise coordinate and Schwarzschild null coordinates
\begin{equation}
  r_* = r + \ln|r-1|,
  \qquad
  u = t-r_*,
  \qquad
  v = t+r_*.
  \label{eq:tortoise}
\end{equation}
In Region~I ($U<0$, $V>0$) one may then take
\begin{equation}
  U=-e^{-u/2},
  \qquad
  V=e^{v/2},
  \label{eq:UVdef}
\end{equation}
so that one has the following global implicit relation valid on all four Kruskal sectors:
\begin{equation}
  UV = -(r-1)e^r,
  \qquad
  r(U,V)=1+W(-UV/e),
  \label{eq:KSrel}
\end{equation}
where $W$ refers to principal Lambert function. In Region~I the Schwarzschild time is recovered from the ratio relation
\[
  t = \frac12(v+u)=\ln\!\left(-\frac{V}{U}\right).
\]
The metric in these coordinates is
\begin{equation}
  \extd s^2 = -\frac{4}{r}e^{-r}\,\extd U\,\extd V
  + r^2\bigl(\extd\theta^2+\sin^2\!\theta\,\extd\phi^2\bigr),
  \label{eq:metricUV}
\end{equation}
with $g_{UV}=g_{VU}=-\frac{2}{r}e^{-r}$, so that
$\sqrt{-g}=2re^{-r}\sin\theta$.
Unlike the Schwarzschild metric \eqref{eq:metrictr}, this one is $C^\infty$-regular at, e.g. the future horizon $U=0$ corresponding to $r=1$.

The non-zero Christoffel symbols of the metric~\eqref{eq:metricUV} are given explicitly as,
\begin{equation}
\label{Christoffel}
\begin{aligned}
  \Gamma^U_{UU} &= \frac{r+1}{r^2}Ve^{-r},
  &\quad
  \Gamma^U_{\theta\theta} &= -\frac{Ur}{2},
  &\quad
  \Gamma^U_{\phi\phi} &= -\frac{Ur}{2}\sin^2\!\theta,
  \\
  \Gamma^V_{VV} &= \frac{r+1}{r^2}Ue^{-r},
  &\quad
  \Gamma^V_{\theta\theta} &= -\frac{Vr}{2},
  &\quad
  \Gamma^V_{\phi\phi} &= -\frac{Vr}{2}\sin^2\!\theta,
  \\
  \Gamma^\theta_{U\theta} &= -\frac{e^{-r}}{r^2}V,
  &\quad
  \Gamma^\theta_{V\theta} &= -\frac{e^{-r}}{r^2}U,
  &\quad
  \Gamma^\theta_{\phi\phi} &= -\sin\theta\cos\theta,
  \\
  \Gamma^\phi_{U\phi} &= -\frac{e^{-r}}{r^2}V,
  &\quad
  \Gamma^\phi_{V\phi} &= -\frac{e^{-r}}{r^2}U,
  &\quad
  \Gamma^\phi_{\theta\phi} &= \cot\theta,
\end{aligned}
\end{equation}
with $r=r(U,V)$ understood throughout.

\subsection{Full 4D geodesic velocity and amplitude-flow equations}
\label{sec:full4d}

The geodesic velocity equation \eqref{eq:flowX}  in local
coordinates $x^\alpha=(U,V,\theta,\phi)$ reads
\begin{equation}
  \dot X^\alpha + X^\beta\partial_\beta X^\alpha
  + \Gamma^\alpha_{\beta\rho}X^\beta X^\rho = 0,
  \qquad \alpha\in\{U,V,\theta,\phi\}.
  \label{eq:velflow4d}
\end{equation}
Using the Christoffel symbols~\eqref{Christoffel}, the four component equations are
\begin{align}
  \dot X^U + X^\beta\partial_\beta X^U
  + \frac{r+1}{r^2}Ve^{-r}(X^U)^2
  - \frac{Ur}{2}\bigl[(X^\theta)^2+\sin^2\!\theta\,(X^\phi)^2\bigr]
  &= 0,
  \label{eq:4dU}\\
  \dot X^V + X^\beta\partial_\beta X^V
  + \frac{r+1}{r^2}Ue^{-r}(X^V)^2
  - \frac{Vr}{2}\bigl[(X^\theta)^2+\sin^2\!\theta\,(X^\phi)^2\bigr]
  &= 0,
  \label{eq:4dV}\\
  \dot X^\theta + X^\beta\partial_\beta X^\theta
  - \frac{2e^{-r}}{r^2}(UX^V+VX^U)X^\theta
  - \sin\theta\cos\theta\,(X^\phi)^2
  &= 0,
  \label{eq:4dTh}\\
  \dot X^\phi + X^\beta\partial_\beta X^\phi
  - \frac{2e^{-r}}{r^2}(UX^V+VX^U)X^\phi
  + 2\cot\theta\,X^\theta X^\phi
  &= 0.
  \label{eq:4dPh}
\end{align}
Moreover, the unit-speed condition $|X|^2=-1$, preserved by~\eqref{eq:flowX}, reads
\begin{equation}
  -\frac{4}{r}e^{-r}X^UX^V + r^2\bigl[(X^\theta)^2+\sin^2\!\theta\,(X^\phi)^2\bigr]
  = -1.
  \label{eq:unitspeed4d}
\end{equation}
The associated amplitude-flow equation for a complex scalar field $\psi(s,x)$ is
\begin{equation}
  \dot\psi+X^\alpha\partial_\alpha\psi+\frac12\,\div(X)\,\psi=0,
  \label{eq:psiPDE4d}
\end{equation}
where the full four-dimensional covariant divergence in Kruskal variables is
\begin{equation}
  \div(X)
  =\partial_\alpha X^\alpha
  +\frac{r-1}{r^2}e^{-r}(VX^U+UX^V)+\cot\theta\,X^\theta .
  \label{eq:divX4d}
\end{equation}
For the geometric-phase analysis below we restrict the above 4D flow system to the axially symmetric case with $\theta$ angular-mode wave-packets.

\subsection{Axial reduction of the full 4D system}
\label{sec:reduction}

We now specialize the full 4D flow equations to the axially symmetric case that will then be used in the geometric-phase calculation in the later sections. 
\[
  X^\phi\equiv0,\qquad \partial_\phi\equiv0
\]
for all fields. This is the natural setting for the angular-mode initial data $e^{ik(\theta-\theta_0)}$ used below. Setting $X^\phi=0$ in~\eqref{eq:4dPh} is consistent (remains true during evolution), while the remaining three equations (\ref{eq:4dU})-(\ref{eq:4dTh}) reduce to
\begin{align}
  \dot X^U + X^\mu\partial_\mu X^U
  + \frac{r+1}{r^2}Ve^{-r}(X^U)^2 - \frac{Ur}{2}(X^\theta)^2 &= 0,
  \label{eq:geodU}\\
  \dot X^V + X^\mu\partial_\mu X^V
  + \frac{r+1}{r^2}Ue^{-r}(X^V)^2 - \frac{Vr}{2}(X^\theta)^2 &= 0,
  \label{eq:geodV}\\
  \dot X^\theta + X^\mu\partial_\mu X^\theta
  - \frac{2e^{-r}}{r^2}(UX^V+VX^U)X^\theta &= 0,
  \label{eq:geodTh}
\end{align}
where, from now on
\begin{equation}
  x^\mu=(U,V,\theta),
  \qquad
  X=X^\mu\partial_\mu,
  \qquad
  \mu,\nu\in\{U,V,\theta\}.
  \label{eq:xmu}
\end{equation}
This is the notation used throughout the rest of the paper. The reduced unit-speed condition is
\begin{equation}
  -\frac{4}{r}e^{-r}X^UX^V + r^2(X^\theta)^2 = -1.
  \label{eq:unitspeed3d}
\end{equation}
The reduced amplitude-flow equation is
\begin{equation}
  \dot\psi + X^\mu\del_\mu\psi
  + \frac{1}{2}\,\div(X)\,\psi = 0,
  \label{eq:psiPDE}
\end{equation}
with reduced covariant divergence
\begin{equation}
  \div(X) = \del_\mu X^\mu
  + \frac{r-1}{r^2}e^{-r}(VX^U+UX^V) + \cot\theta\,X^\theta.
  \label{eq:divX}
\end{equation}

\subsection{Covariant continuity equation and conserved mass}
The amplitude flow equation can be reformulated into a continuity equation for density
\[
  \rho(s,x):=|\psi(s,x)|^2.
\]
To see this, multiply \eqref{eq:psiPDE} by $\psi^*$, add the complex conjugate, and use the reality of $X$ to get
\begin{equation}
  \dot\rho + \nabla_\mu(\rho X^\mu)=0.
  \label{continuity}
\end{equation}
This naturally leads to a notion of conserved total mass (provided the boundary flux vanishes) for evolving wave-function, given by
\begin{equation}
  N(s):=\int_M \rho(s,x)\,\extd\mu\ ,
  \label{eq:mass}
\end{equation}
where the measure is over the full $4$D spacetime manifold $M$, so in our case
\begin{equation}
  \extd\mu = 4\pi re^{-r}\sin\theta\,\extd U\,\extd V\,\extd\theta.
  \label{eq:measure}
\end{equation}
This is the basic conservation law behind the centroid diagnostics used later.

\subsection{Ehrenfest identities for centroid transport}\label{sec:ehrenfest}

For any smooth scalar observable $f=f(s,x)$, we can define its expectation value as
\begin{equation}
  \langle f\rangle(s):=\frac{1}{N}\int_M f(s,x)\,\rho(s,x)\,\extd\mu.
  \label{expectf}
\end{equation}
Differentiating with respect to $s$, using \eqref{continuity}, and integrating by parts on the manifold gives the following Ehrenfest identity
\begin{equation}
  \frac{\extd}{\extd s}\langle f\rangle
  =\bigl\langle \dot f + X(f)\bigr\rangle= \bigl\langle \frac{D}{Ds}\Big|_X f\bigr\rangle
  \label{ehrenMaster}
\end{equation}
in terms of the convective derivative. The centroid transport equations follow immediately by taking $f=U,V,\theta$:
\begin{equation}
  \frac{\extd\langle U\rangle}{\extd s}=\langle X^U\rangle,\qquad
  \frac{\extd\langle V\rangle}{\extd s}=\langle X^V\rangle,\qquad
  \frac{\extd\langle\theta\rangle}{\extd s}=\langle X^\theta\rangle.
  \label{ehren1}
\end{equation}
These are the precise mathematical justification for using
the packet centroid in numerical diagnostics below, as they follow the averaged geodesic velocity
field.

Applying \eqref{ehrenMaster} to the $s$-dependent observable $f=X^\mu$ then gives
\begin{equation}
  \frac{\extd^2\langle x^\mu\rangle}{\extd s^2}=\langle \frac{D}{Ds}\Big|_X X^\mu\rangle,
  \label{ehren2}
\end{equation}
as the second-order Ehrenfest relation. Using the geodesic velocity equation
\eqref{eq:velflow4d} componentwise,
\[
   \frac{D}{Ds}\Big|_X X^\mu=\dot X^\mu+X^\nu\partial_\nu X^\mu=-\Gamma^\mu_{\nu\sigma}X^\nu X^\sigma,
\]
so we have  equivalently,
\begin{equation}
  \frac{\extd^2\langle x^\mu\rangle}{\extd s^2}
  =-\bigl\langle \Gamma^\mu_{\nu\sigma}X^\nu X^\sigma\bigr\rangle\approx- \Gamma^\mu_{\nu\sigma}(\bar x)\langle X^\nu\rangle\langle X^\sigma\rangle. 
  \label{eq:ehren2geom}
\end{equation}
Here, $\bar x=\langle x\rangle$ is the centroid location, and the approximation is in the narrow-packet limit, where the density $\rho$ is sharply concentrated near the centroid $\bar{x}$ and variations of both Christoffel and velocity terms are negligible across packets. Thus, in this particle-like regime, the centroid itself approximately obeys the ordinary geodesic equation with $s$ proper time. 

 \section{Framework for geometric Aharonov--Bohm phase }
\label{sec:phaseframework}

We will work in the reduced system (\ref{sec:reduction}) with $U,V,\theta$ as coordinates. The first step is the construction of a suitable geodesic velocity field $X$. Then we discuss the wave-packet form of $\psi$ as well as the notion of backward integral curves ending at chosen observation points. This leads to a phase formula which will be used in the numerical analysis in Section~\ref{sec:numerics}.

\subsection{Construction of the geodesic velocity field}
\label{sec:Xinit}

The initial geodesic velocity field at $s=0$ is constructed by starting from a 2D field in the $(U,V)$ sector and then lifting it into the current 3D setting. The 2D fields with components $\widetilde X = \big( \widetilde X^U(U,V), \widetilde X^V(U,V) \big)$, are required to satisfy two conditions as in~\cite{KumarMajid2025},
\begin{equation}
  \partial_U\widetilde X^U + \partial_V\widetilde X^V
  + \frac{r-1}{r^2}e^{-r}\bigl(V\widetilde X^U + U\widetilde X^V\bigr) = 0,\quad  -\frac{4}{r}e^{-r}\widetilde X^U\widetilde X^V = -1.
\label{eq:seedconds}
\end{equation}
The first is the 2D covariant divergence-free condition $\div(\widetilde X)=0$ and the
second is the Lorentzian unit-speed condition $|\widetilde X|^2=-1$ in the $(U,V)$-sector. Solving the second for $\widetilde X^V$ gives
\begin{equation}
  \widetilde X^V = \frac{re^r}{4\widetilde X^U},
  \label{eq:seedXV}
\end{equation}
reducing the first to a single second-order PDE for $\widetilde X^U$, solved
numerically with Dirichlet data $\widetilde X^U\vert_{U_{\min}}=5$ on the inflow
face $U_{\min}=-4$, providing an ingoing flow. We then choose the initial 3D data at $s=0$ as
\begin{equation}
\label{eq:Xinit}
\begin{split}
  X^U(0,U,V,\theta)
    &= \widetilde X^U(U,V)\,\sqrt{1+r(U,V)^2\sin^2\!\theta},\\
  X^V(0,U,V,\theta)
    &= \widetilde X^V(U,V)\,\sqrt{1+r(U,V)^2\sin^2\!\theta},\\
  X^\theta(0,U,V,\theta)
    &= \sin\theta,
\end{split}
\end{equation}
which, owing to \eqref{eq:seedconds}, satisfies the 3D unit-speed condition \eqref{eq:unitspeed3d} and restricts to $\widetilde{X}$ at the poles $\theta=0,\pi$. These 3D fields also preserve the $U,V$ symmetry of the system. Furthermore, the choice $X^\theta(0)=\sin\theta$ ensures that the field vanishes at the poles of the unit-sphere. 

\begin{figure}[t]
\centering
\includegraphics[width=\textwidth]{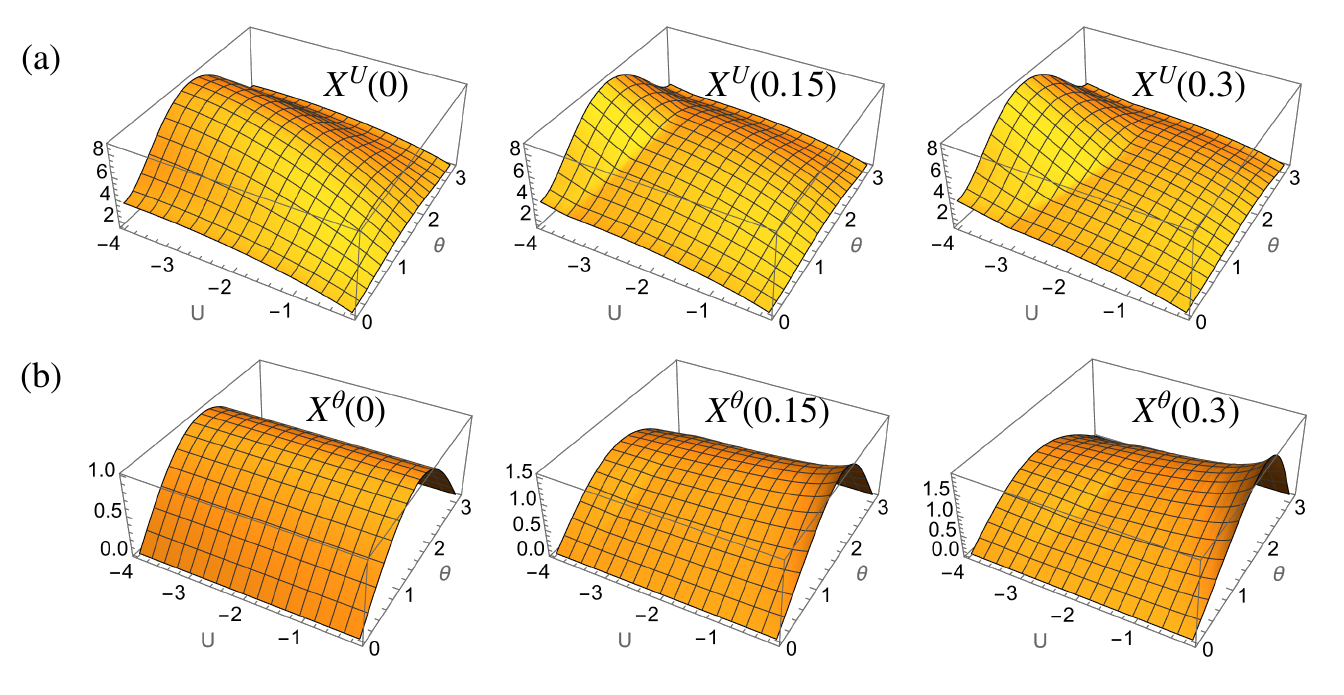}
\caption{Plots of velocity field at $s=0,0.15,0.3$ shown for components (a) $X^U(s)$ and (b) $X^\theta(s)$.}
\label{fig:velocityfield}
\end{figure}

Once the initial velocity fields have been specified as above, the full field $X(s,U,V,\theta)$ is obtained for $s>0$ by numerically evolving the geodesic velocity equations \eqref{eq:geodU}--\eqref{eq:geodTh}. Throughout the rest of the paper we work in Region I on the domain
\begin{equation}
  U\in[-4,-0.1],\qquad V\in[0.1,6],\qquad \theta\in[0.05,\pi-0.05],
  \label{eq:domains}
\end{equation}
where the safe angular cut-off $0.05$ keeps the amplitude-flow equation away from the coordinate singularities at the two poles. We solve the PDE system using NDSolve of Mathematica using the MethodOfLines for $s$ with FiniteElements for $(U,V,\theta)$ and grid-cell resolution ${\rm res}={\rm MaxCellMeasure}=0.005$, while fixing $X=X(0)$ at the $U_{\min}=-4$ boundary. The angular component $X^\theta$, that appears in the geometric phase formula \eqref{phaseformula}, is positive throughout the specified part of Region I. We illustrate the resulting fields $X^U(s)$ and $X^\theta(s)$ thus obtained with $V=3$ slices at $s=0,0.15,0.3$ in Figure~\ref{fig:velocityfield}. The plots for $X^V(s)$ are similar to that of $X^U(s)$ so we refrain from showing those here.

\subsection{Wave-packets carrying the phase}
\label{sec:wave-packets}

We next consider initial data for the numerical runs for $\psi_k$  as a 3D Gaussian wave-packet centred at $(U_0,V_0,\theta_0)$ with the same width $\sigma$ along all three coordinates,
\begin{equation}
  \psi_k(0,U,V,\theta)
  = G(U,V,\theta)\,e^{ik(\theta-\theta_0)},
  \label{eq:initdata}
\end{equation}
where the Gaussian envelope is $k$-independent and given by
\begin{equation}
  G(U,V,\theta) = \exp\!\Bigl[-\frac{(U-U_0)^2+(V-V_0)^2+(\theta-\theta_0)^2}{2\sigma^2}\Bigr].
  \label{eq:Gdef}
\end{equation}
For the mode number $k$, we focus the numerical analysis below in three of them:
\begin{center}
\begin{tabular}{ll}
\toprule
$k=0$ & reference envelope mode, used to isolate amplitude transport,\\
$k=2$ & main interference mode together with $k=0$ mode,\\
$k=4$ & higher-mode confirmation of exact $k$-linearity.\\
\bottomrule
\end{tabular}
\end{center}
The case \(k=1\) can also be computed, but the shift is more visible for \(k=2\) and higher.
As the wave-packet \eqref{eq:initdata} evolves according to the amplitude flow \eqref{eq:psiPDE}, it is transported and deformed by the flow. A natural way to track the packet motion is through the centroid, namely the averaged position weighted by $|\psi|^2$, whose evolution is governed by the Ehrenfest identities discussed in Section~\ref{sec:ehrenfest}. This is the usual wave-packet viewpoint, in which a localised packet is compared with a centre or reference trajectory and the deformation of nearby support \cite{Littlejohn1986}. For the phase law, however, the more precise object is the backward integral curve of the vector field $X$ ending at the chosen observation point at time $s$. The observation point may be taken to be the centroid, but the backward integral curve need not retrace the centroid path exactly; the mismatch gets smaller with narrow packets.

\subsection{Backward integral curves and exact solution}\label{sec:back}

For a fixed point $x^\mu=(U,V,\theta)$ at flow time $s$, we define $\gamma^\mu(\tau)$ to be the
backward integral curve ending at $x^\mu$ at time $s$, characterised by
\begin{equation}
  \dot\gamma^\mu(\tau)=X^\mu(\tau,\gamma(\tau)),
  \qquad
  \gamma^\mu(s)=x^\mu,\quad \tau\le s.
  \label{eq:IC}
\end{equation}
Now for any smooth function $f(\tau,x)$ we have,
\begin{align*}
  \frac{\extd}{\extd\tau}f(\tau,\gamma(\tau))
  &= \frac{\partial f}{\partial\tau}(\tau,\gamma(\tau))
    + \dot\gamma^\mu(\tau)\,\partial_\mu f(\tau,\gamma(\tau)) \\
  &= \frac{\partial f}{\partial\tau}(\tau,\gamma(\tau))
    + X^\mu(\tau,\gamma(\tau))\,\partial_\mu f(\tau,\gamma(\tau)).
  \label{eq:chainrule}
\end{align*}
Applying this to $f=\psi$ and using \eqref{eq:psiPDE}, we obtain
\begin{align*}
  \frac{\extd}{\extd\tau}\psi(\tau,\gamma(\tau))
  &= \bigl(\partial_\tau + X^\mu\partial_\mu\bigr)\psi(\tau,\gamma(\tau))\\
  &= -\tfrac12\,\div(X)(\tau,\gamma(\tau))\,\psi(\tau,\gamma(\tau)).
\end{align*}
Integrating this scalar ODE along $\gamma$ gives the exact solution
\begin{equation}
  \psi(s,x)=\psi(0,\gamma(0))
  \exp\!\left(-\frac12\int_0^s\div(X)(\tau,\gamma(\tau))\,\extd\tau\right).
  \label{eq:exactsol}
\end{equation}
Since $\div(X)$ is real, the exponential factor is real and positive, hence the phase of $\psi(s,x)$ is exactly the initial phase evaluated at the foot
$\gamma(0)$. We label the foot of the backward integral curve by
\[
x_f:=\gamma(0)=(U_f,V_f,\theta_f).
\]

Before moving further, it is important to distinguish two distinct angular quantities entering our analysis:
\begin{itemize}
\item $\theta_0$ is the \emph{fixed angular reference} built into the initial data \eqref{eq:initdata}. It is the angular centre of the initial Gaussian blob and as such is a constant.
\item $\theta_f=\theta_f(s,x)$ is the \emph{foot-point angle}, a dynamical quantity obtained by back-tracing the integral curve $\gamma$ which at flow time $s$ lands at $x=(U,V,\theta)$.
\end{itemize}
With this distinction established, we can now integrate the $\theta$-component of \eqref{eq:IC},
\begin{equation}
  \theta_f(s,x)=\theta-\int_0^s X^\theta(\tau,\gamma(\tau))\,d\tau.
  \label{thetaf}
\end{equation}
Hence the net transported angular displacement relative to the fixed initial reference angle $\theta_0$ is
\begin{equation}
  \Delta\theta(s,x):=\theta_f(s,x)-\theta_0
  =(\theta-\theta_0)-\int_0^s X^\theta(\tau,\gamma(\tau))\,d\tau.
  \label{delthetaf}
\end{equation}

\subsection{Phase formula and exact \texorpdfstring{$k$}{k}-linearity.}
\label{sec:klaw}

For the initial data \eqref{eq:initdata} with $G>0$ real and independent of $k$, the exact solution \eqref{eq:exactsol} gives
\[
  \psi_k(s,x)=G(\gamma(0))\,e^{ik(\theta_f-\theta_0)}\,F(s,x),
\]
where $\gamma(0)=x_f$ is the foot-point noted above and
\[
  F(s,x)=\exp\!\left(-\frac12\int_0^s\div(X)(\tau,\gamma(\tau))\,d\tau\right)>0.
\]
Note that the modulus is independent of $k$:
\begin{equation}
  |\psi_k(s,x)| = A(s,x):=G(\gamma(0))\,F(s,x).
  \label{modulus}
\end{equation}
The full phase is $k(\theta_f-\theta_0)=k\Delta\theta(s,x)$, so using \eqref{delthetaf} it splits as
\begin{equation}
  \arg\psi_k(s,x)=k(\theta-\theta_0)+\varphi_k(s,x),
  \label{phasefull}
\end{equation}
where the reduced phase encoding the observable shift during the flow is 
\begin{equation}
  \varphi_k(s,x)
  := \arg\!\bigl(\psi_k(s,x)e^{-ik(\theta-\theta_0)}\bigr)
  = -k\int_0^s X^\theta(\tau,\gamma(\tau))\, \extd\tau.
  \label{phaseformula}
\end{equation}
We will be using this precise phase quantity in Section~\ref{sec:numerics} to compare against the observable phase-shift of the numerical $\psi(s,x)$ data obtained through the flow equation \eqref{eq:psiPDE}. An important point to note here is the exact $k$-linearity relation in the above phase formula; any deviation noted in Section \ref{sec:numerics} should be treated as purely numerical error. Finally, one may note that Gaussian envelope $G$ \eqref{eq:Gdef} depends on $\theta$ alongside $(U,V)$, but this dependence enters only through $|\psi(s,x)|$ and hence does not play any role in the phase-shift analysis below.

%-------------------------------------------------------------------
\subsection{Interference observable}
\label{sec:obs}
%-------------------------------------------------------------------
Having established the geometric setting for the wave-packet evolution and the exact phase law, we now look at how this will be detected through  observation of the intensity $|\psi_k|^2$. We consider coherent two-mode superpositions
\[
  \psi_{0k}:=\psi_0+\psi_k=\psi_0(1+e^{ki\Delta\theta}).\]
Using \eqref{modulus}, we have  the pure cross-term 
\begin{equation}
  |\psi_{0k}|^2-|\psi_0|^2-|\psi_k|^2 = 2|\psi_0||\psi_k|\cos(k\Delta\theta).
  \label{crossTerm}
\end{equation}

We then use \eqref{crossTerm} to obtain the observable phase shift from the normalised interference cross-term as
\begin{equation}
\begin{aligned}
  \bar R_k(s,\theta)
  &=\frac{N_k(s,\theta)}{D_k(s,\theta)},\\
  N_k(s,\theta)
  &=
 2 \int\bigl(|\psi_{0k}|^2-|\psi_0|^2-|\psi_k|^2\bigr)
     re^{-r}\,\extd U\,\extd V,\\
  D_k(s,\theta)
  &=4
  \int  |\psi_0||\psi_k|
   re^{-r}\,\extd U\,\extd V .
\end{aligned}
  \label{eq:Rbar}
\end{equation}
In the ideal equal-envelope regime $|\psi_0|=|\psi_k|$ that we are interested in, this \eqref{eq:Rbar} reduces to 
\[ \bar R_k=\cos(k\Delta\theta)\]
since this is also the local ratio at each point of $U,V$. Integrating as we do is better adapted to a packet drifting in the null coordinates $(U,V)$, since the numerator and denominator are first integrated over the packet support before taking the ratio, removing some of the numerical noise. However, as the $\psi_k$ are evolved separately as a test of data integrity, $\bar R_k$ won't exactly have this form due to numerical errors in the numerical evolution. We also have an issue when  \(D_k\) is small as will see later. 

The ideal peak position $\bar R_k=1$ occurs at $\theta_\ast(s)$, satisfying $\Delta\theta(s,\theta_\ast)=2\pi n/k$ for integer $n$. Note that the packet used in the PDE runs in Section~\ref{sec:numerics} occupies only a small angular window around the original centre: $\theta_0\pm 2\sigma$ with $\sigma=0.15$ means $4\sigma_\theta=0.6$ is only about $38\%$ of the half-period $\pi/2\approx1.57$ of $\cos(2\Delta\theta)$. For $k=4$, the corresponding half-period is $\pi/4\approx0.785$, so the same packet spans a larger fraction of the local fringe leading to a narrower principal profile. In both cases the observed phase shift is read from the principal interference feature
\begin{equation}
  \varphi_k^{\bar R}(s) := -\bigl(\theta_\ast^{(k)}(s) -\theta_0\bigr)=\frac{\varphi_k(s)}{k}.
  \label{eq:RbarPhase}
\end{equation}
Here \(\varphi_k(s)\) is the reduced phase from \eqref{phaseformula}, while \(\varphi_k^{\bar R}\) is the normalised intensity-level phase read from the interference profile.

\subsection{Summary of Aharonov--Bohm analogy}

We can now summarize the precise sense in which the mechanism is AB-like. The analogy is not gauge-theoretic: there is no electromagnetic vector potential and no ordinary solenoid-type holonomy. Rather, the analogy is at the level of phase accumulation and interference readout. The angular mode number \(k\) plays the role of a charge, the velocity component \(X^\theta\) drives the accumulated phase, and the observable signature is a displacement of the interference profile. The correspondences are:
\begin{center}
\begin{tabular}{ll}
\toprule
AB effect & Geodesic velocity field\\
\midrule
Charge $q$ & Angular mode number $k$\\
Vector potential $A_\mu$ & Velocity component $X^\theta$\\
Line integral $\oint A\cdot dl$ & $\int_0^s X^\theta(\tau,\gamma(\tau))\,d\tau$\\
Inaccessible solenoid & Black-hole interior\\
Observable: interference shift & $\bar R_k(\theta,s)$ peak shift\\
\bottomrule
\end{tabular}
\end{center}
Thus the effect is AB-like at the level of interference structure, but its phase is generated by geodesic transport in the amplitude-flow equation rather than by a gauge potential.

%-------------------------------------------------------------------

%-------------------------------------------------------------------
\section{Numerical setup and results}
\label{sec:numerics}
%-------------------------------------------------------------------

We now test the phase law numerically using the axially reduced system of Section \ref{sec:reduction}, the reduced phase formula of Section \ref{sec:klaw}, and the interference based phase extraction of Section \ref{sec:obs}. The numerical work has three parts: first we check that the packet remains localised while being transported by the computed geodesic velocity field, followed by the computation of the reduced phase directly from the solved wave-functions and finally the extraction of the corresponding intensity-level phase from the fitted interference profile centre of the \(UV\)-integrated ratio \(\bar R_k\).

The geodesic velocity field \(X(s,U,V,\theta)\) constructed in Section~\ref{sec:Xinit} is used as a fixed transport background for the \(\psi\)-evolution. However, before proceeding, we first sample the velocity field $X$ and its divergence ${\rm div}(X)$ on a coarser grid of size $0.05$ to speed up the computation, without affecting the overall velocity profiles. Using this interpolation, we solve the amplitude flow equation \eqref{eq:psiPDE} for the angular wave-packet data (\ref{eq:initdata})-(\ref{eq:Gdef}) for five different cases, $\psi_k, \psi_{0k}$ with $k=0,2,4$, as explained in Section \ref{sec:obs}. Like before in solving for $X$, we use Mathematica's NDSolve with the MethodOfLines for \(s\), with FiniteElements for \((U,V,\theta)\) and the same spatial resolution scale \({\rm MaxCellMeasure}=0.005\). A Dirichlet condition is imposed only on the inflow boundary $U_{min}=-4$. Although we ran the evolution of these $\psi$ to get raw data till \(s=0.4\), we present the phase comparisons below for \(0.05\leq s\leq0.30\), since at later times boundary effects become visible.

\begin{figure}[!t]
\centering
\includegraphics[width=\textwidth]{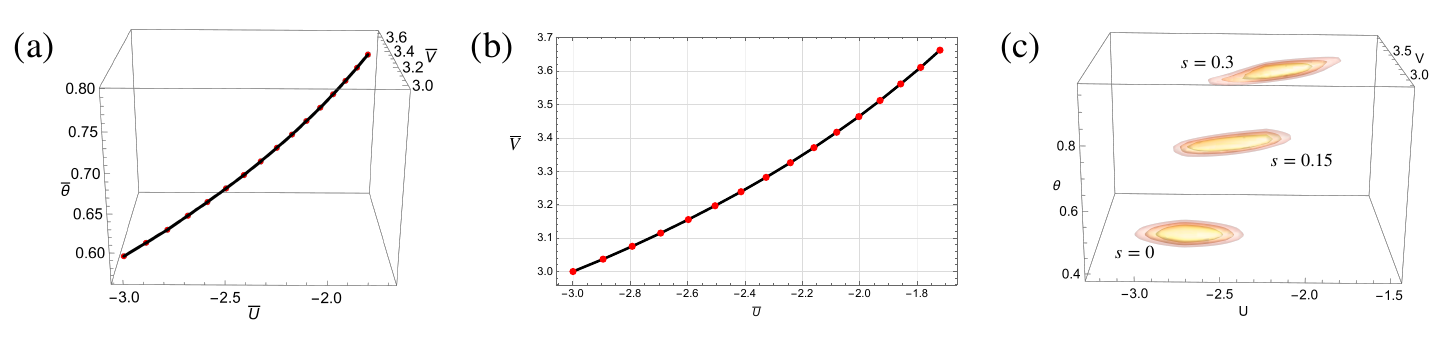}
\caption{Plots showing wave-packet transport with (a) 3D centroid trajectory, (b) 2D projection on centroid trajectory on \((U,V)\) plane and (c) density $\rho(s)$ level-surfaces at $s=0,0.15,0.3$.}
\label{fig:packetmotion}
\end{figure}

\begin{figure}[!t]
\centering
\includegraphics[width=\textwidth]{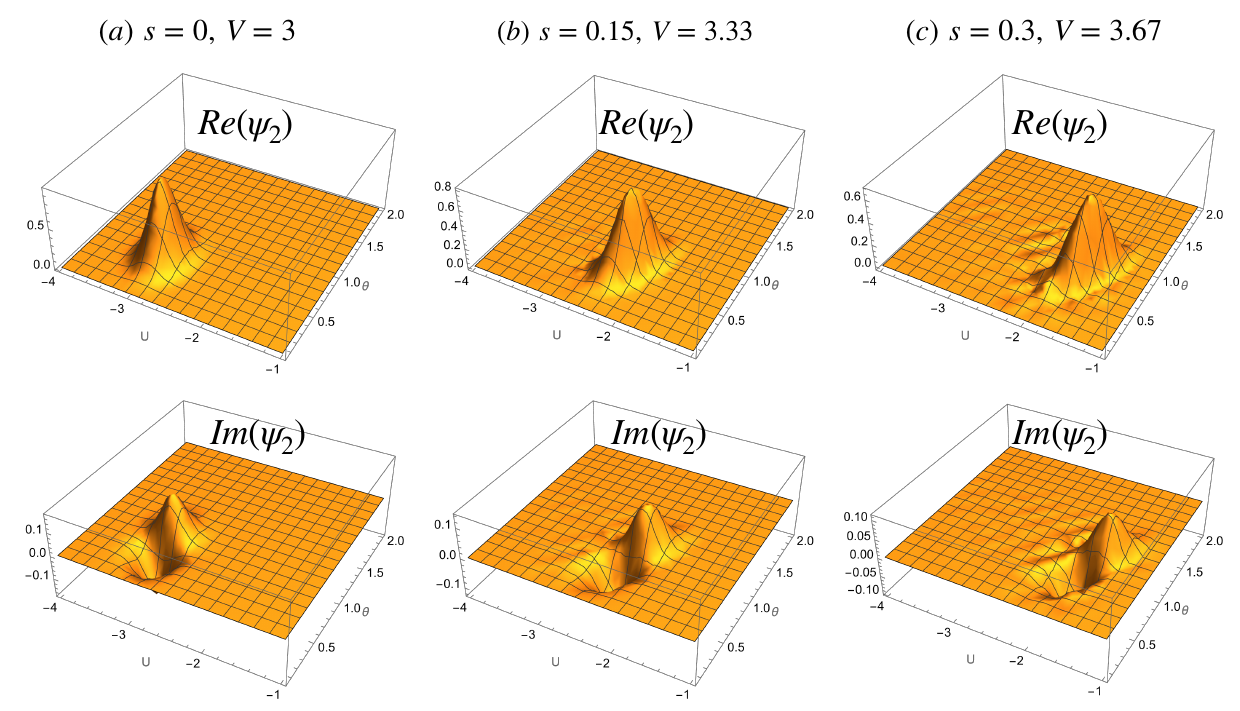}
\caption{Plots showing evolution of wave-function $\psi_2(s)$ at $s=0,0.15,0.3$ and sliced at  corresponding centroid values of $V$. }
\label{fig:psiEvolution}
\end{figure}

\subsection{Packet motion and numerical checks}

We first analyze the packet motion itself in Figure~\ref{fig:packetmotion} showing the transported centroid,
\begin{equation}
   \bar x(s)=(\bar U(s),\bar V(s),\bar\theta(s)),
   \label{eq:centroid}
\end{equation}
its projection to the \((U,V)\)-plane and density level-surfaces for the \(k=0\) packet with level-values $(0.6,0.7,0.8)$. The centroid moves towards the \(U=0\) horizon with increasing \(V\) and \(\theta\), in agreement with the positive angular component \(X^\theta\) of the velocity field. We show the complex amplitude of $\psi_2$ itself in Figure~\ref{fig:psiEvolution} on slices at the centroid value \(V=\bar V(s)\)  as it evolves in $s$, this being where the real and imaginary parts remain localised. The ripples visible at later times (i.e., column (c)) are numerical noise from the evolution. The corresponding plots for $\psi_0$ (always real) and $\psi_4$ are similar.

To estimate the numerical accuracy we monitor the mass \(N(s)\) \eqref{eq:mass} and find that for single-modes $k=0,2,4$ these remain conserved to within \(10^{-3}\). We also directly verify that the independently evolved interference fields $\psi_{0k}$ remain linearly related in the correct manner, in that the residual
\[
  \mathcal E_{0k}(s) := \int \bigl|\psi_{0k}(s)-\psi_0(s)-\psi_k(s)\bigr|^2\,d\mu, \qquad k=2,4,
\]
stays within \(10^{-6}\)--\(10^{-5}\) during the evolution. Similarly the amplitudes of individual wave-functions $\psi_k$ match to similar high accuracy, thereby showing phase-independent amplitude evolution \eqref{eq:exactsol}. Thus the evolutions are stable and free from issues such as non-linearity of interference, mass drift, etc meaning that larger numerical discrepancies below in phase analysis can be attributed to finite-resolution or other numerical effects.

\subsection{Direct reduced phase extraction}

\begin{figure}[!t]
\centering
\includegraphics[width=\textwidth]{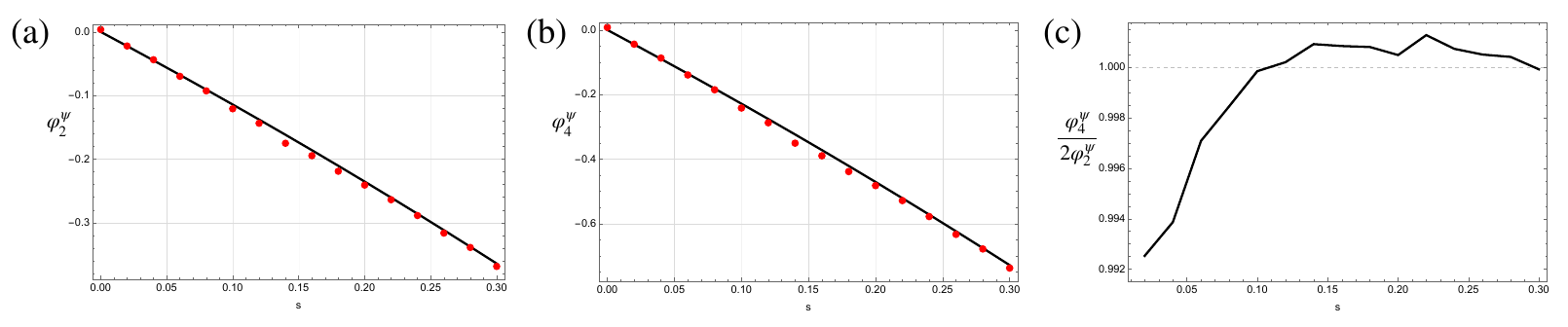}
\caption{Comparison plots of direct reduced phases $\varphi^{\psi}_k(s)$ against the flow prediction $\varphi_k(s)$ obtained from the velocity field component $X^\theta(s)$ for (a) $k=2$ and (b) $k=4$. The normalised ratio $\frac{\varphi^{\psi}_4}{2\varphi^{\psi}_2}(s)$ in (c) stays close to 1 as an integrity check of linearity of the phase law.}
\label{fig:directphase}
\end{figure}

We first evaluate the reduced phase directly from the above obtained wave-functions by evaluating \eqref{phaseformula} along the transported centroid \eqref{eq:centroid}:
\[
  \varphi_k^\psi(s)
  =
  \arg\!\left[
    \psi_k(s,\bar x(s))e^{-ik(\bar\theta(s)-\theta_0)}
  \right].
\]
This is compared with the flow prediction $\varphi_k(s)$ given by \eqref{eq:phase-law} along the corresponding backward integral curve, evaluated numerically (point-wise evaluations at $\delta s=0.02$ intervals via a trapezoidal rule along the centroid, followed by interpolation). Figure~\ref{fig:directphase} compares the \(k=2\) and \(k=4\) extracted phases with their flow predictions, and also shows the numerical ratio \(\varphi_4^\psi/(2\varphi_2^\psi)\). This ratio tests the exact \(k\)-linearity relation %\eqref{eq:phaselinearity} 
by showing that this remains close to 1 during the evolution (slightly low initial values are due to an onset effect for a phase-difference starting from zero). Therefore, this direct wave-function-level extraction offers a sharp numerical validation of the geometric phase law.

\subsection{Phase extraction from intensity profiles}

\begin{figure}[!b]
\centering
\includegraphics[width=\textwidth]{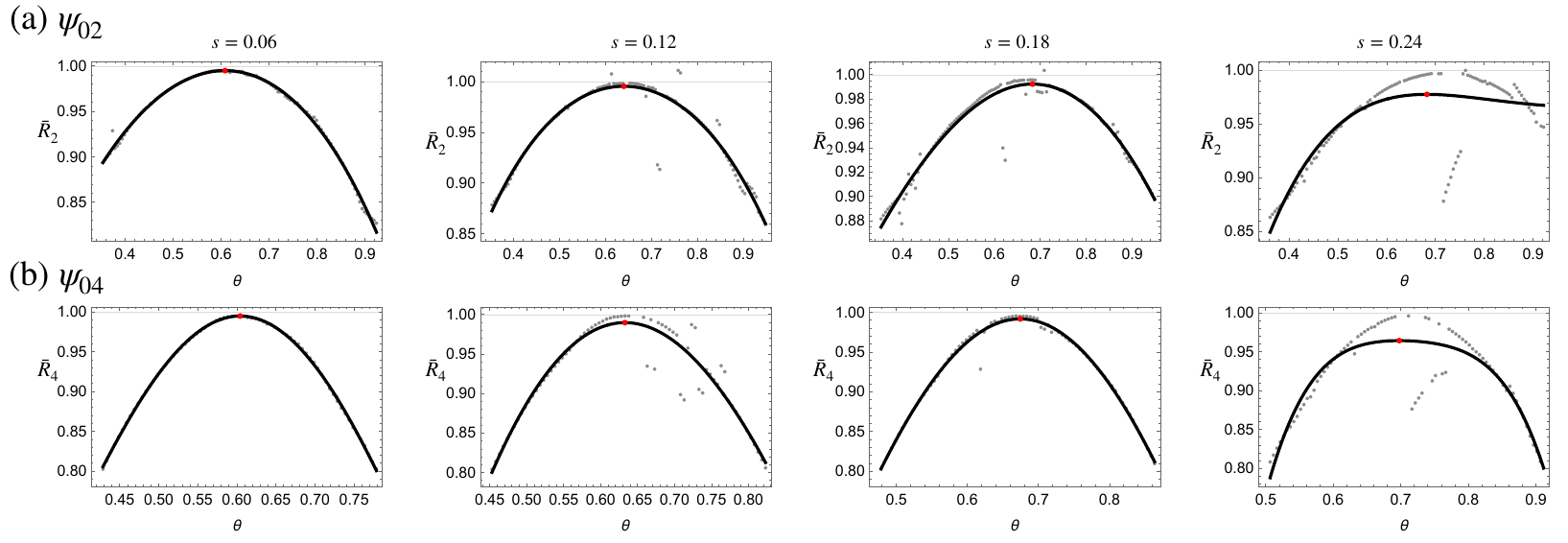}
\caption{Plots of \(\bar R_k\) sample data, fitted low degree curve and peak location (red dot) at $s=0.06,0.12,0.18,0.24$ for the interference profiles (a) $\psi_{02}$, and (b) $\psi_{04}$.}
\label{fig:rbarfitexamples}
\end{figure}

As discussed in Section \ref{sec:obs}, the ratio of $(U,V)$-integrated cross-term $N_k$ and overlap-term $D_k$ provides a more physically meaningful way of extracting the phase, since intensities are closer to laboratory readouts than the complex wave-function itself. However, this ratio is also more delicate computationally because of the denominator overlap integrals that can sometimes become too small and produce artefacts. Keeping this in mind, we obtain this ratio \(\bar R_k\) \eqref{eq:Rbar} through post-processing of the solved fields $\psi_k$ where sampling is done on a grid of size \(\delta s=0.02\) and \(\delta\theta=0.005\). At each \(s\), the raw profile is sampled in a packet-tracking angular window \(\bar\theta(s)\pm 0.35\) to capture large enough dynamic intervals and record the raw data \((s,\theta,\bar R_k,D_k)\).

\begin{figure}[!t]
\centering
\includegraphics[width=\textwidth]{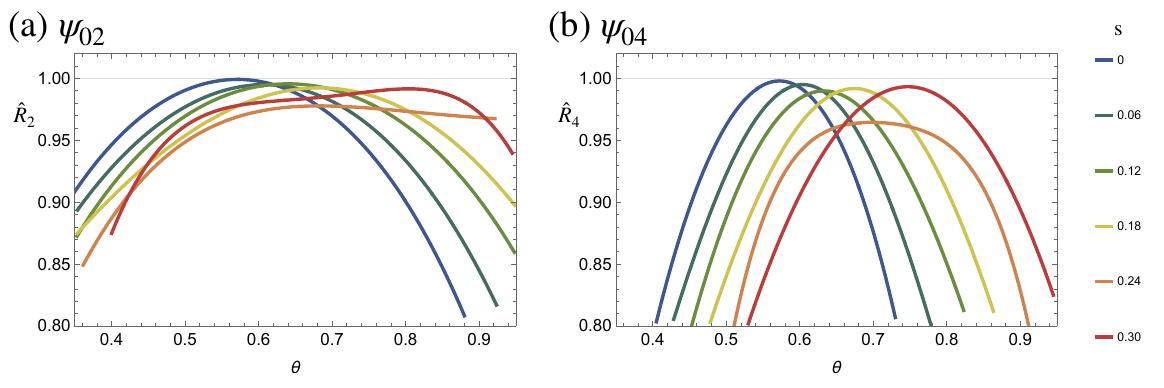}
\caption{Evolution of the fitted interference profiles \(\widehat R_k(s,\theta)\) for (a) \(k=2\) and (b) \(k=4\). We show reconstructed fitted-curve snapshots at $s=0,0.06,0.12,0.18,0.24,0.3$.}
\label{fig:rbarfitprofiles}
\end{figure}

\begin{figure}[!t]
\centering
\includegraphics[width=\textwidth]{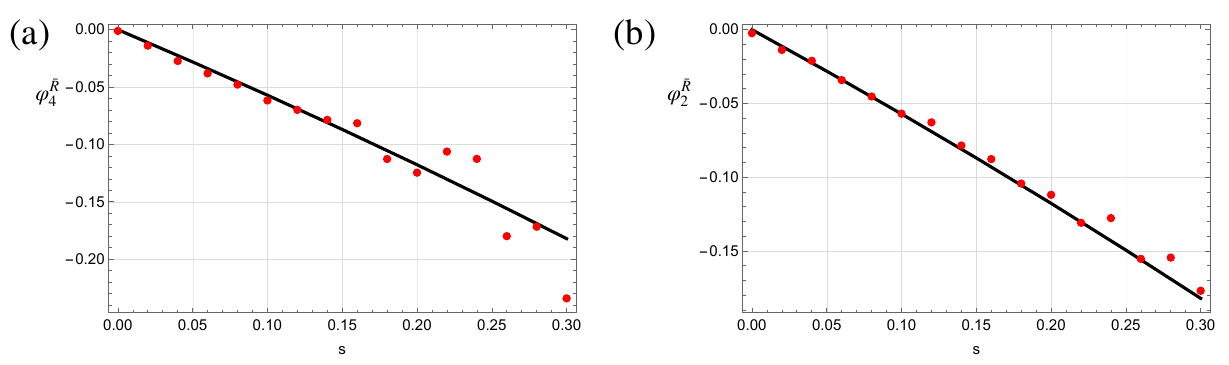}
\caption{Fitted intensity-level phase \(\varphi_k^{\bar R}\) compared with the
normalised flow prediction \(\varphi_k/k\).}
\label{fig:rbarphase}
\end{figure}

We then apply some natural filtering on this data to remove the far-tail of the profiles as well as bad sample points. More specifically, for each \(s\)-slice we work in a fixed angular band \(0.35\le\theta\le0.95\), discarding points with \(D_k\le10^{-3}\), and applying a range-filter \(0.80\le \bar R_k\le1.03\). We then fit a low-degree curve, 
\[
\widehat R_k(s,\theta) = a_0 + a_1\theta + a_2\theta^2 + a_3\theta^3 + a_4\theta^4,
\]
of degree at most four on the filtered data. The fitted peak \(\theta_\ast^{(k)}(s)\) defines the normalised intensity-level phase \(\varphi_k^{\bar R}(s)\) as in \eqref{eq:RbarPhase}. Representative sampled profiles and fitted curves are shown in Figure~\ref{fig:rbarfitexamples}, the fitted-profile evolution is shown in Figure~\ref{fig:rbarfitprofiles}, and the resulting phase comparison with \(\varphi_k(s)/k\) is shown in Figure~\ref{fig:rbarphase}. We have also shown the sampled points alongside the fitted curve, showing the presence of a few numerically sensitive outlier points as well as more systematic late-time distortions (peak flattening), especially after about \(s\simeq0.22\), due to erratic denominator $D_k$ values. This explains why a fitted centre is more accurate than trying to find maximum from the raw data itself.

The \(k=2\) profile in Figure~\ref{fig:rbarfitprofiles} (a) is broader compared to the \(k=4\) profile in part (b) because \(\cos(k\Delta\theta)\) varies more rapidly with \(\theta\) for the latter. Compared to the previous direct phase comparison,  the ratio-based extraction in this section is more sensitive  to numerical errors in the evolution of the different $\psi_0,\psi_k,\psi_{0k}$. In particular, resulting small or uneven values of \(D_k\) and late-time flattening can move the fitted maxima. The fitted-centre procedure reduces this sensitivity, and the narrower \(k=4\) profiles give slightly cleaner peak locations as evident from Figure \ref{fig:rbarphase}(b).

\FloatBarrier
%-------------------------------------------------------------------
\section{Conclusion}
%-------------------------------------------------------------------

In this work we have extended the time-radial analysis of geodesic flows around a black hole in \cite{KumarMajid2025} to include the angular coordinate
$\theta$ (the angle from the vertical axis), while assuming symmetry (i.e. $\phi$-independence) about the axis. Each dimension added makes the numerical integrations less tractable so only allowed variation in $\theta$. This was enough, however, to demonstrate a novel geometric AB effect under the hypothesis of complex amplitudes underlying matter densities on spacetime. We first set up a geodesic velocity field (\ref{eq:Xinit}) which has a certain $U-V$ profile at the poles but in between the poles `bulges' to largest values at the equator as can be seen in Figure~\ref{fig:velocityfield}. 

We then looked at initial wave-packet wave-functions $\psi_k$ with phase corresponding to frequency $k$ in the $\theta$ direction and density centred as a `Gaussian bump' at some $(U_0,V_0,\theta_0)$. In our case small $\theta_0$, so including the axial symmetry we have a kind of ring around the vertical axis. We then evolved everything to later $s$. The centroid angle increases as the `ring' approaches the black hole and we saw how the phases evolve and how their interference evolves. In principle one could continue evolving past the black-hole to centroid values beyond $\pi/2$ but one would need to overcome some numerical issues arising from close to the horizon.  The theoretical analysis led to a formula (\ref{phaseformula}) for the phase shift of wave packet which we interpreted in terms of the interference of a linear combination with $\psi_0$. The theoretical values were then computed for the case of interest as well as directly for the numerical evolutions with a good match well above numerical noise levels. Although not an actual experiment, the analysis shows a theoretical prediction backed by numerical modelling. 

The theoretical underpinning in Section~\ref{sec:pre} shown that the classical quantum geodesic  flow equations in fact arise naturally from the Klein-Gordon flow or Stueckelberg proper time quantum mechanics, which in turn is the Schr\"odinger picture of the generally covariant quantum mechanics recently introduced in \cite{BegMa:gen}. We also saw that the convective derivative $\div(X)$ of a geodesic velocity field found in \cite{BegMa:cur} involving the Ricci curvature can be viewed as a version of the Raychaudhuri equations in relativistic fluid mechanics.

Overall, we find that the hypothesis that in some contexts the density in relativistic fluid mechanics could be replaced by an underlying spacetime wave-function amplitude leads to new interference effects (of an AB-type). Moreover, the framework we use is now shown to arise naturally from a recent formulation of generally covariant quantum mechanics \cite{BegMa:gen} introduced in both Heisenberg form and in Schr\"odinger form where it corresponds to a Klein-Gordon flow. Both have been studied around a black-hole\cite{BegMa:gen, KumarMajid2025} and in FLRW backgrounds\cite{BegMa:flrw}. As a result, while matters of physical interpretation need to be further explored due to the novel nature of this approach, it appears that there is a natural theoretical basis and the possibility to compute examples, both of which should be developed further. In this respect, other background spacetimes could be studied. A useful next step would be to recast the geodesic family studied in \cite{KumarMajid2025} into standard congruence language, including comoving/Lemaître-type frames, expansion, shear, vorticity and caustics in the sense of Section~\ref{sec:ray}.  This should clarify how the present amplitude flow mechanism relates to the ADM lapse-shift description of the same black-hole geometry. Also to be developed further is the adaptation of ideas from relativistic fluid mechanics to the amplitude setting as again suggested by the present work. These are some directions for further work.

\section*{Declarations}

\textbf{Acknowledgements.}
KK was supported by DFG project grant 515782239. SM was supported by Leverhulme Trust project grant RPG-2024-177. The numerical computations involved use of high performance computing facilities at QMUL.

\textbf{Data availability.}
No experimental data were created in this work. All supporting code and numerical data used to generate the figures are publicly available at
\cite{Mathematica}.

\textbf{Conflict of interest.}
The authors have no competing interests to declare that are relevant to the content of this article.


\begin{thebibliography}{30}

\bibitem{AharonovBohm}
Y.~Aharonov and D.~Bohm,
Significance of electromagnetic potentials in the quantum theory,
\textit{Phys.\ Rev.}\ \textbf{115} (1959) 485--491.

\bibitem{Anandan}
J.~Anandan,
Gravitational and rotational effects in quantum interference,
\textit{Phys.\ Rev.\ D} \textbf{15} (1977) 1448--1457.

\bibitem{Beg:geo} 
E.J. Beggs, 
Noncommutative geodesics and the KSGNS construction, 
J. Geom. Phys. 158 (2020) 103851.

\bibitem{BegMa}
E.J.~Beggs and S.~Majid,
\emph{Quantum Riemannian Geometry},
Grundlehren der mathematischen Wissenschaften \textbf{355}, Springer (2020).

\bibitem{BegMa:cur}
E.J.~Beggs and S.~Majid,
Quantum geodesic flows and curvature,
\textit{Lett.\ Math.\ Phys.}\ \textbf{113} (2023) 73.

\bibitem{BegMa:geo}
E.J.~Beggs and S.~Majid,
Quantum geodesics in quantum mechanics,
\textit{J.\ Math.\ Phys.}\ \textbf{65} (2024) 012101.

\bibitem{BegMa:gen}
E.J.~Beggs and S.~Majid,
Generally covariant quantum mechanics,
\textit{Lett.\ Math.\ Phys.}\ \textbf{116} (2026) 9.

\bibitem{BegMa:flrw}
E.J. Beggs and S. Majid, Klein--Gordon flow on FLRW spacetimes,
arXiv:2501.18295 [gr-qc] (2025).

\bibitem{Berry}
M.V.~Berry,
Quantal phase factors accompanying adiabatic changes,
\textit{Proc.\ Roy.\ Soc.\ A} \textbf{392} (1984) 45--57.

\bibitem{ColellaOverhauserWerner}
R.~Colella, A.W.~Overhauser and S.~A.~Werner,
Observation of gravitationally induced quantum interference,
\textit{Phys.\ Rev.\ Lett.}\ \textbf{34} (1975) 1472--1474.

\bibitem{Coo} J. Cooke, Proper-time formulation of quantum mechanics, Physical Review, Vol 166 (1968) 1293-1298

\bibitem{Dowker}
J.S.~Dowker,
A gravitational Aharonov--Bohm effect,
\textit{Nuovo Cimento B} \textbf{52} (1967) 129--135.

\bibitem{Hohensee}
M.A.~Hohensee, B.~Estey, P.~Hamilton, A.~Zeilinger and H.~M\"uller,
Force-free gravitational redshift: proposed gravitational Aharonov--Bohm experiment,
\textit{Phys.\ Rev.\ Lett.}\ \textbf{108} (2012) 230404.

\bibitem{HorPir} L.P. Horwitz and C. Piron, Relativistic dynamics, Helvetica Physica Acta Vol. 46 (1973)

\bibitem{Jusufi}
K. Jusufi, A. Yasser, E. Battista and N. Inan,
Signatures of modified gravity from the gravitational Aharonov--Bohm effect,
\textit{arXiv}:2502.07613 [gr-qc] (2025).

\bibitem{Mathematica}
K.~Kumar and S.~Majid,
Mathematica codes for Geometric Aharonov--Bohm phase effect around a black hole,
\textit{Notebook Archive} \href{https://notebookarchive.org/2026-05-c16632a}{https://notebookarchive.org/2026-05-c16632a}.

\bibitem{KumarMajid2025}
K.~Kumar and S.~Majid,
Geodesic flows on a black-hole background,
\textit{arXiv}:2603.03222 [gr-qc] (2026).

\bibitem{LawrenceLeiterSzamosi}
J.K.~Lawrence, D.~Leiter and G.~Szamosi,
A gravitational ``Aharonov--Bohm'' effect,
\textit{Nuovo Cimento B} \textbf{17} (1973) 113--121.

\bibitem{Littlejohn1986}
R.G.~Littlejohn,
The semiclassical evolution of wave packets,
\textit{Phys.\ Rep.}\ \textbf{138} (1986) 193--291.

\bibitem{Overstreet}
C.~Overstreet, P.~Asenbaum, J.~Curti, M.~Kim, and M.A.~Kasevich,
Observation of a gravitational Aharonov--Bohm effect,
\textit{Science} \textbf{375} (2022) 226--229.

\bibitem{Papini}
G.~Papini,
Particle wave-functions in weak gravitational fields,
\textit{Nuovo Cimento B} \textbf{52} (1967) 136--141.

\bibitem{Stachel}
J.~Stachel,
Globally stationary but locally static space-times: a gravitational analog of
the Aharonov--Bohm effect,
\textit{Phys.\ Rev.\ D} \textbf{26} (1982) 1281--1290.

\bibitem{StachelPlebanski}
J.~Stachel and J.~Plebański,
Classical particles with spin. I. The WKBJ approximation,
\textit{J.\ Math.\ Phys.}\ \textbf{18} (1977) 2368--2374.

\bibitem{Stodolsky}
L.~Stodolsky,
Matter and light wave interferometry in gravitational fields,
\textit{Gen.\ Relativ.\ Gravit.}\ \textbf{11} (1979) 391--405.

\bibitem{Stu} E.C.G. Stueckelberg, The mechanics of the material point in Relativity theory and in quantum theory, Helvetica Physica Acta, vol. 15 (1942) 23--37.



\end{thebibliography}
\end{document}